\documentclass[aps,pra,reprint,twocolumn,superscriptaddress,showkeys,a4paper]{revtex4-2}
\usepackage{amsfonts}
\usepackage{amssymb}
\usepackage{mathrsfs}
\usepackage[intlimits]{amsmath}
\usepackage{bm, bbm}
\usepackage[dvips]{graphicx}
\usepackage[T1]{fontenc}
\usepackage[latin2]{inputenc}
\usepackage{amsmath}

\usepackage{latexsym}

\usepackage{color}

\def\dt{{\rm d}\,}

\newcommand{\ket}[1]{| #1 \rangle}
\newcommand{\bra}[1]{\langle #1 |}

\def\duzomniejsze{<\kern-.7mm<}
\def\duzowieksze{>\kern-.7mm>}

\def\textbf#1{{\bf #1}}
\def\be{\begin{equation}}
\def\ee{\end{equation}}
\def\ben{\begin{eqnarray}}
\def\een{\end{eqnarray}}
 \def\beqa{\begin{eqnarray}}
\def\eeqa{\end{eqnarray}}
\def\eea{\end{array}}
\def\bea{\begin{array}}
\newcommand{\bei}{\begin{itemize}}
\newcommand{\eei}{\end{itemize}}
\newcommand{\bee}{\begin{enumerate}}
\newcommand{\eee}{\end{enumerate}}

\def\1{\openone}

\def\tr{{\rm Tr}}

\def\>{\rangle}
\def\<{\langle}

\def\dt#1{{{\kern -.0mm\rm d}}#1\,}
\def\squareforqed{\hbox{\rlap{$\sqcap$}$\sqcup$}}
\def\qed{\ifmmode\squareforqed\else{\unskip\nobreak\hfil
\penalty50\hskip1em\null\nobreak\hfil\squareforqed
\parfillskip=0pt\finalhyphendemerits=0\endgraf}\fi}

\newtheorem{lemma}{Lemma}
\newtheorem{theorem}[lemma]{Theorem}
\newtheorem{main result}[lemma]{Main result}
\newtheorem{proposition}[lemma]{Proposition}
\newtheorem{definition}{Definition}

\newtheorem{fact}[lemma]{Fact}

\def\bep{\begin{proposition}}
\def\eep{\end{proposition}}
\def\bel{\begin{lemma}}
\def\eel{\end{lemma}}

\def\bet{\begin{theorem}}
\def\eet{\end{theorem}}
\def\bed{\begin{definition}}
\def\eed{\end{definition}}
\def\bef{\begin{fact}}
\def\eef{\end{fact}}

\begin{document}

\title{Decoherence and objectivity in higher spin environments}

\author{M. Kici\'nski }
\affiliation{Warsaw University of Technology, Physics Department, 00-662 Warsaw, Poland}
\author{J. K. Korbicz}
\email{jkorbicz@cft.edu.pl}
\affiliation{Center for Theoretical Physics, Polish Academy of Sciences, 02-668 Warsaw, Poland}
\date{\today}

\begin{abstract}
We analyze decoherence and objectivization processes in spin-spin models for arbitrary spins. We first derive the most general analytic form of the decoherence factor in the measurement limit, where the interaction Hamiltonian dominates the rest. We then analyze 
thermal environments and derive exact, analytic formulas for both the decoherence factor and the state fidelity of post-interaction environment states. This allows to analyze the objectivization process of the state of the central spin during the interaction. We do so using, so called, Spectrum Broadcast Structures (SBS), which are specific multipartite quantum states encoding a certain operation notion of objectivity. We analyze analytically (for short times) and numerically how higher spin influences the efficiency of decoherence and objectivization processes. As expected, the higher the spin of the environment, the more efficient decoherence and objectivization become. This work is a generalization of previous studies, limited to spin-$1/2$ systems only, and we hope will be useful in future objectivity experiments.  
\end{abstract}


\maketitle

\section{Introduction}
In recent years it has been realized that decoherence \cite{decoh} alone is not enough to explain the apparent classicality of the macroscopic world, even if we accept the probabilistic character of decoherence output \cite{ZurekNature, PRA}. The missing part is the objective, i.e. observer independent, character of the classical world. At a first glance, this objectivity seems to be at odds with the quantum theory as in quantum mechanics every act of observation inevitably disturbs the state of the observed object. However, if quantum theory is to be at the foundation of our understanding of Nature, it should explain somehow how the objective character of classical quantities emerges during quantum-to-classical transition. Such mechanisms of objectivization have been indeed proposed. The first, and most popular, is known under the name of quantum Darwinism \cite{ZurekNature}. It recognizes that most of the observations in the macroscopic world are done indirectly, with broadly understood environment mediating them and thus being a source of valuable information about the system of interest.  It then states that information about the system (e.g. position) becomes objective if it is: i) replicated during the evolution in many copies in systems' environment and ii) these copies are available for a read-out without disturbance. The formal part of the idea, based on a information-theoretic condition, has been however criticized as not necessarily delivering what is promised \cite{PRA, Le} and a new approach, based directly on quantum state structure has been proposed in \cite{PRL, PRA} under the name of Spectrum Broadcast Structures (SBS) (see \cite{solo} for a detailed analysis of different approaches). It states that an information about a system becomes objective in the above operational sense if the system and an observed part of its environment are in a specific state called SBS \cite{PRA, solo}:
\be\label{sbs0}
\varrho_{S:E_{obs}}=\sum_i p_i |i\rangle\langle i| \otimes \varrho_i\otimes\cdots\otimes\varrho_i, \text{ where } \varrho_i\perp\varrho_{i'\ne i}.
\ee
Here $|i\rangle$ are so called pointer states, i.e. stable states into which the system decoheres, $p_i$ are the initial pointer probabilities, and $\varrho_i$ are states of the environment fragments, having their supports orthogonal for different indices $i$ (denoted by $\perp$). The information that is being made objective via SBS is the index $i$, identifying the state of the central system. Condition (\ref{sbs0}) is a ramification of the standard decoherence condition, as it limits not only the form of the reduced state of the system alone, but how it is correlated to portions of the environment. It is also stronger than the original quantum Darwinism condition \cite{solo, PRL, Mironowicz_PRL} (it can be relaxed a bit to so called strong quantum Darwinism \cite{Le, solo}, allowing for some correlations within the environment,  but it will not be necessary in this work). 
 
SBS states were found in a series of quantum open system models \cite{Tuziemski qbm, Lewenstein, QED, gravity, measurements, GPT, Kasia} including spin-spin models \cite{Mironowicz_PRL, Mironowicz PRA, NV}. The latter models, just like the majority of other spin-spin studies (see e.g. \cite{Zurek spins, Mauro}), were restricted to spin-$1/2$ systems, with the restriction being most limiting for the environment, affecting its efficiency as an information medium. There are however systems of practical interest possessing higher spins, for example single-molecule magnets (see e.g. \cite{magnets}), and it would be beneficial to generalize the previous studies to such systems. Here, we perform this generalization and study objectivity processes in arbitrary spin-$j$ models. We consider a central spin $j_S$ coupled to a collection of spin-$j$  systems, constituting its environment. Decoherence in such systems was analyzed before e.g. in \cite{Hamdouni}, but objectivity so far not. We complement the previous decoherence studies deriving the most general expression for the decoherence factor in the quantum measurement limit. Since every spin-$j$ system can be simulated by a totally symmetric subspace of $2j$ spins-$1/2$,  it is natural to expect that increasing the spin will be equivalent to taking a larger amount of spin-$1/2$ systems. We show that this is indeed so and in general increasing the spin of the environment helps to approach SBS state more efficiently. This means that in practical simulations of objectivization processes, which has started to appear recently  \cite{Ciampini, Chen, ZurekJelezko}, higher spins allow to reduce the number of environmental spins that have to be controlled. We derive and analyze exact expressions for both the decoherence factor and the state fidelity as functions of $j$ in a realistic case of thermal environment, which in the context of SBS was studied only very recently in \cite{NV} for spin-$1/2$ environment. We thus generalize the original SBS analysis of spin-spin model \cite{Mironowicz_PRL} in two directions: i) by assuming an arbitrary spin $j$ and ii) by adding a non-trivial Hamiltonian for the environment.  

To introduce the basic setup, let us first assume a simplified situation of the quantum measurement limit, in which the system-environment interaction: 
\begin{equation}\label{H}
H_{SE}=S_z\otimes \sum_{k=1}^N g_k S_z^{(k)}
\end{equation}
dominates over all other energies. We assume the central spin to be described by the spin-$j_S$ representation of $SU(2)$, while all the environmental spins coming from the spin-$j$ space. Here, $S_z$ and $S_z^{(k)}$ are the $z$-axis spin operators for the central system and the $k$-th environment respectively, acting in the respective spin-$j_S$ and spin-$j$ spaces, and $g_k$ are the coupling constants, controlling the strength of the interaction. The spin operators $S_i$ satisfy the usual commutation relations:
\be
[S_i,S_j]=i\epsilon_{ijk} S_k.
\ee
The evolution due to (\ref{H}) is easily solved by diagonalizing $S_z=\sum_{m=-j_S}^{j_S}m\ket{j_S;m} \bra{j_S;m}$:
\be\label{H1}
H_{SE}=\sum_{m=-j_S}^{j_S}\ket{m} \bra{m} \otimes \sum_{k=1}^N m g_k S_z^{(k)}.
\ee
[In what follows we drop the $j_S$ index for clarity; magnetic numbers $m,m'$ will always pertain to the central spin and range $-j_S$ to $+j_S$.] We then find that $U_{S:E}=e^{-itH}$ reads:
\be\label{USE}
U_{S:E}=\sum_{m=-j_S}^{j_S}\ket{m} \bra{m} \otimes \bigotimes_{k=1}^N U_m^{(k)}(t),
\ee
where
\be\label{Um}
U_m^{(k)}(t)\equiv e^{-itm g_k S_z^{(k)}}
\ee
We assume a totally uncorrelated initial state as we will be interested in the system-environment correlations buildup:
\be\label{rhoSE0}
\varrho_{S:E}(t=0)=\sigma_{0S}\otimes \bigotimes_{k=1}^N \varrho_{0k}.
\ee
After evolving for time $t$ and discarding a part of the environment assumed to be unobserved, we are left with the so called partially traced state:
\begin{eqnarray}
&&\varrho_{S:E_{obs}}(t)=\tr_{E_{unobs}} \left[ U_{S:E}\varrho_{S:E}(0)U_{S:E}^\dagger\right]\\
&&=\sum_{m=-j_S}^{+j_S} \alpha_{m} \ket{m} \bra{m}\otimes \bigotimes_{k\in E_{obs}} \varrho^{(k)}_m(t)\label{mama1}\\
&&+ \sum_{m\ne m'} \alpha_{mm'} \Gamma_{mm'}(t) \ket{m} \bra{m'} \otimes \bigotimes_{k\in E_{obs}} U_m^{(k)} \varrho_{0k} U_{m'}^{(k)\dagger},\label{mama2}
\end{eqnarray}
where 
\be\label{Gamma}
\Gamma_{mm'}(t)\equiv \prod_{k\in E_{unobs}}\tr\left[\varrho_{0k}U_{m'}^{(k)\dagger}(t) U_m^{(k)}(t)\right]
\ee
is the decoherence factor due to the unobserved environment $E_{unobs}$, $\alpha_{mm'}\equiv \langle m| \sigma_{0S} |m'\rangle$ with $\alpha_{m}\equiv\alpha_{mm}$ being the initial pointer probabilities, and
\be\label{rhom}
\varrho_m^{(k)}(t)\equiv U_m^{(k)} \varrho_{0k} U_{m}^{(k)\dagger}.
\ee
The partially traced state $\varrho_{S:E_{obs}}(t)$  will be our main object of study. In particular, we are interested under which conditions it comes close to the nearest SBS state, so that the spin value $m$ of the central spin becomes objective (in the SBS sense).
Intuitively, from (\ref{mama1},\ref{mama2}) it should happen when: i) the coherent part (\ref{mama2}) disappears and ii) the states $\varrho_m^{(k)}(t)$ become distinguishable for $m\ne m'$. 
In \cite{Mironowicz_PRL} this was formalized and it was proven that for pure dephasing Hamiltonians of the type (\ref{H1}) the approach to the nearest SBS state is controlled by the expression:
\ben
			&& \frac{1}{2} \min ||\varrho_{S:E_{obs}}(t) - \varrho_{SBS}||_{tr} \leq  \sum_{m\ne m'} |\alpha_{mm'}| |\Gamma_{mm'}(t)|\nonumber\\
			&&+\sum_{m\ne m'} \sqrt{\alpha_m\alpha_{m'}}\sum_{k\in E_{obs}} F \left( \varrho_{m}^{(k)}(t), \varrho_{m'}^{(k)}(t) \right),\label{approach}\\
\een
where
\be\label{fid}
F(\varrho, \sigma)\equiv\tr\sqrt{\sqrt\varrho\sigma\sqrt\varrho},
\ee
is the state fidelity, appearing here as a measure of distinguishability \cite{Fuchs}. We recall that $\varrho, \sigma$ have orthogonal supports and hence are one-shot perfect distinguishable if and only if $F(\sigma, \varrho)=0$.
In many situations the interaction is too weak, or the information capacity of the states $ \varrho_{m}^{(k)}(t)$ to small, to have an appreciable distinguishability of $ \varrho_{m}^{(k)}(t)$ for different $m$'s. For example the central spin $j_S$ may be larger than the environmental spins $j$, thus the dimensionality of a single environment too low to have a chance to faithfully encode the state of the central spin. To remedy such cases, we will use a simple coarse-graining trick \cite{PRL}: We group the observed environment
$E_{obs}$ into groups called macrofractions, each consisting of some number $\mu N$, $0<\mu<1$ of the environments. The state of a macrofraction in the studied scenario is a simple product (cf. (\ref{mama1})), $\varrho_m^{mac}(t)=\bigotimes_{k\in mac} \varrho_m^{(k)}(t)$
and hence the fidelity is a product too:
\be
F \left( \varrho_{m}^{mac}(t), \varrho_{m'}^{mac}(t) \right)=\prod_{k\in mac}F \left( \varrho_{m}^{(k)}(t), \varrho_{m'}^{(k)}(t) \right). \label{Fmac}
\ee
As a product of numbers smaller than unity, it has now a better chance to vanish. More advanced metrological scenarios are possible but for the purpose of this work we will use this simple product trick.

\section{Decoherence from an arbitrary environment}

We first briefly recall the decoherence process in the quantum measurement limit (\ref{H}). Using (\ref{USE}-\ref{rhoSE0}), one easily finds the decoherence factor from a single environment (we neglect the $k$-index for clarity):
\begin{equation}
\gamma_{mm'}(t)=\tr\left[\varrho_0 \exp\left(-\text i gt \Delta m S_z\right)\right], \label{Gamma}
\end{equation}
where $\Delta m \equiv m-m'$. Let us first study the case when the environment is initially in a spin-coherent state \cite{Perelomov}, so that $\varrho_0=\ket{\mathbf n}\bra{\mathbf n}$, where:
$\mathbf n\equiv (\sin\theta\cos\phi, \sin\theta\sin\phi, \cos\theta)$ and 
\begin{equation}\label{spin coh}
\ket{\mathbf n}=\exp\left[-\text i \theta (\sin\phi S_x - \cos\phi S_y)\right]\ket{j;-j} \equiv e^{-\text i \theta \mathbf{r S}}\ket{j;-j}.
\end{equation}
%The reason is to pick a physically interesting but low dimensional subspace in the space of all spin-$j$ states, which is $(2j+1)$-dimensional. 
Inserting (\ref{spin coh}) into (\ref{Gamma}) 
we find:
\be\label{G_pure0}
\gamma^{pure}_{mm'}(t)=\bra{\mathbf n}\exp\left(-\text i gt\Delta m S_z\right)\ket{\mathbf n}.
\ee
Thus $\gamma_{mm'}(t)$ is a generating function for the moments of $S_z$ in the state $\ket{\mathbf n}$. It was calculated explicitly e.g. in Appendix B of \cite{Arecchi} and reads:
\be\label{G_pure}
\gamma^{pure}_{mm'}(t)=\left[\cos\left(\frac{gt \Delta m}{2}\right)+ i \sin\left(\frac{gt \Delta m}{2}\right)\cos\theta\right]^{2j},
\ee
so that
\be\label{G_pure mod}
|\gamma^{pure}_{mm'}(t)|^2=\left[\cos^2\theta+\cos^2\left(\frac{gt \Delta m}{2}\right)\sin^2\theta\right]^{2j}.
\ee
There is a full restoration of the initial coherences, i.e.  $|\gamma^{pure}_{mm'}(t)|^2=1$, for $gt \Delta m=2k\pi$. From the symmetry of the setup -- both the spin coherent states and the coupling to the environment are defined via $S_z$,  $\gamma^{pure}_{mm'}(t)$ does not depend on the azimuth angle $\phi$ (spin coherent states depend on $\phi$ via phase factors $e^{in\phi}$ that cancel in (\ref{G_pure0})). It also follows from the form of  (\ref{G_pure}) that in this simple case of pure environment, the decoherence factor is a $2j$-power of the decoherence factor for spin-$1/2$, so the spin value behaves literary like a particle number. For short times $gt\Delta m\ll 1$ we find a Gaussian decay:
\be
|\gamma^{pure}_{mm'}(t)|\approx \exp\left(-\frac{j}{4} \sin^2\theta g^2 \Delta m^2 t^2\right).
\ee

Let us now consider a general initial state of the environment. It is well known \cite{Perelomov} that each state in the spin-$j$ representation can be diagonally expanded in the spin coherent states, analogously to optical coherent states:
\be\label{rhoP}
\varrho_0=\int d^2\mathbf{n} P(\mathbf n) \ket{\mathbf n}\bra{\mathbf n}.
\ee
Further expanding $P(\mathbf n)$ in spherical harmonics in the usual way $P(\mathbf n)=\sum_{l,\mu} c_{l\mu}Y_{l\mu}(\theta, \phi)$ we obtain a parametrization, given by a non-commutative analog of the Fourier transform:
\be\label{rhoY}
\varrho_0=\sum_{l,\mu} c_{l\mu}\hat Y_{l\mu},
\ee
where
\be
\hat Y_{l\mu}=\int d^2\mathbf{n} Y_{l\mu}(\theta, \phi) \ket{\mathbf n}\bra{\mathbf n}.
\ee
A few remarks are in order. Firstly, we recall \cite{Perelomov} that unlike in the optical case, $\varrho_0$ determines $P(\mathbf n)$ only up to the order of $l=2j$ in the spherical harmonics decomposition. We thus set all terms with $l>2j$ to zero.
The normalization of $\varrho_0$ fixes the $c_{00}$ coefficient to:
\be
c_{00}=\frac{1}{2\sqrt \pi}.
\ee 
Finally, there are also additional conditions on $c_{l\mu}$ that guarantees the positivity of $\varrho_0$. However, these conditions, being a non-commutative analog of the classic Bochner theorem, are complicated and we will not elaborate on them.
Substituting (\ref{rhoY}) into (\ref{Gamma}) we obtain:
\be
\gamma_{mm'}(t)=\sum_{l\mu} c_{l\mu} \int d^2\mathbf{n} Y_{l\mu}(\theta, \phi) \gamma^{pure}_{mm'}(t),
\ee
where $\gamma^{pure}_{mm'}(t)$ is the pure-state decoherence factor (\ref{G_pure}). Since $\gamma^{pure}_{mm'}(t)$ does not depend on $\phi$, we can set $c_{l\mu}=0$ for $\mu\ne 0$ and call $c_{l0}$ simply $c_l$. Using the binomial decomposition of
(\ref{G_pure}), we are then left with integrals of Legendre polynomials with power function. This can be easily done and finally gives:
\begin{eqnarray}
&&\gamma_{mm'}(t)= \nonumber \\ 
&&\sqrt\pi\sum_{l=0}^{2j}c_l\sqrt{(2l+1)}\sum_{k=l}^{2j} {2j\choose{k}}c(t)^{2j-k}[is(t)]^k I_{lk},\label{gamma_general}
\end{eqnarray}
where $c(t)\equiv\cos(gt \Delta m/2)$, $s(t)\equiv \sin(gt \Delta m/2)$, and
\be
I_{lk}=\begin{cases} 0, & \text{for}\ (-1)^{kl}=-1\\
\frac{(-1)^n\Gamma(r+\frac{1}{2})}{\Gamma(n+r+\frac{3}{2})}, & \text{for}\ l=2n, k=2r\\
\frac{(-1)^n\Gamma(r+\frac{3}{2})}{\Gamma(n+r+\frac{5}{2})}, & \text{for}\ l=2n+1, k=2r+1
\end{cases}
\ee
with $\Gamma(r)$ being the Euler gamma function. This is the most general expression for the single-spin decoherence factor. The full decoherence factor due to the unobserved environment is given according to (\ref{Gamma}) by the product of
single-spin factors (\ref{gamma_general}) with different $k$. The price to pay for the full generality is the dependence on $2j$ parameters $c_l$, which makes even the single-spin expression (\ref{gamma_general}) difficult to study
even numerically for moderate spins $j$. Another drawback is that calculation of the fidelity function (\ref{fid}), needed to study the approach to SBS per (\ref{approach}), seems rather hopeless in this full generality. We therefore choose a different, yet more realistic strategy and study thermal, rather than arbitrary environments.

\section{Thermal environment}

Consider a more realistic Hamiltonian, including dynamics of the environment:
\begin{equation}\label{Hth}
H=2S_z\otimes \sum_{k=1}^N g_k S_z^{(k)}-2\sum_{k=1}^N\Omega_k S_x^{(k)},
\end{equation}
where $\Omega_k$ is the tunneling energy and we have introduced the factor two for the later convenience. Diagonalizing again $S_z=\sum_{m=-j_S}^{j_S}m \ket m \bra m$ we can rewrite (\ref{Hth}) as a pure dephasing Hamiltonian:
\be
H=\sum_{m=-j_S}^{j_S}\ket m \bra m\otimes \sum_{k=1}^N H^{(k)}_m,
\ee
where 
\be\label{Hm}
H^{(k)}_m\equiv2\left[ mg_k S_z-\Omega_k S_x\right].
\ee
The dynamics is then easily found formally:
\be
U=\sum_{m=-j_S}^{j_S}\ket m \bra m\otimes \bigotimes_{k=1}^N U_m^{(k)}, 
\ee
where now
\be
U_m^{(k)}=\exp\left[-2it\left(mg_k S_z^{(k)}-\Omega_k S_x^{(k)}\right)\right].\label{Um'}
\ee
Let us assume the environment is initially in the thermal state with respect to its free Hamiltonian $H_E=-2\sum_{k=1}^N\Omega_k S_x^{(k)}$ with some inverse temperature $\beta$:
\ben
&&\varrho_{0E}= \bigotimes_{k=1}^N\frac{1}{Z_k}e^{-2\beta \Omega_k S_x^{(k)}}, \label{rho0}\\
&& Z_k=\tr \left(e^{-2\beta \Omega_k S_x^{(k)}}\right)=\frac{\sinh\left[(2j+1)\beta\Omega_k\right]}{\sinh\left(\beta\Omega_k\right)}.
\een
The decoherence factor due to a single environment then reads:
\be\label{gmm'}
\gamma_{mm'}(t)=\frac{1}{Z}\tr\left[e^{-2\beta \Omega S_x} U_{m'}^\dagger U_m\right].
\ee
It can be calculated using e.g. the method of \cite{Arecchi}, based on the fact that $U_m$ are representations of $SU(2)$ group elements and relations among group elements hold irrespectively of the representation. We thus first set $j=1/2$, so that all the matrices are just $2\times 2$ and the calculations can be done explicitly. The matrix:
\be
M\equiv  e^{-2\beta \Omega S_x} U_{m'}^\dagger U_m, 
\ee
which trace we have to calculate, is now from $SL(2,\mathbf C)$ and can be brought to the upper triangular form by an $SU(2)$ transformation, preserving the trace and the determinant. The diagonal values are just the eigenvalues, satisfying:
\be\label{Viet}
\lambda_++\lambda_-=\tr M,\quad \lambda_+\lambda_-=\det M=1.
\ee
Then choosing for definiteness $\lambda\equiv\lambda_+$, the trace can be written as 
\be
\tr M=\sum_{l=-1/2}^{+1/2}\lambda^{2l}=\frac{\lambda^2-\lambda^{-2}}{\lambda-\lambda^{-1}}.
\ee
The key observation is that fro arbitrary spin $j$, the upper triangular form of $M$ has powers of $\lambda$ on the diagonal and it is enough to extend the above sum to the range from $l=-j$ to $l=+j$. We note that the magnetic numbers $m,m'$ pertain to the central system and are independent of the environmental magnetic number $l$. The statistical sum $Z$ can be recovered by the same summation formula by setting $m'=m$. We denote the corresponding eigenvalue by $\lambda_0$. Using the partial sum expression for the geometric series, we obtain a closed formula for the decoherence factor \cite{Arecchi}:
\be\label{G_full}
\gamma_{mm'}(t) =\frac{\left(\lambda^{2j+1}-\lambda^{-2j-1}\right)/\left( \lambda-\lambda^{-1}\right)}{\left(\lambda_0^{2j+1}-\lambda_0^{-2j-1}\right)/\left( \lambda_0-\lambda_0^{-1}\right)}.
\ee
The eigenvalue $\lambda$ can be easily calculated from (\ref{Viet}):
\be
\lambda=\kappa+\sqrt{\kappa^2-1},
\ee
where:
\begin{eqnarray}
&& \kappa\equiv \gamma_0\cosh(\beta\Omega)-i\gamma_x\sinh(\beta\Omega),\label{kappa}\\
&&\gamma_0\equiv \cos(\omega_m t)\cos(\omega_{m'} t)+\nonumber\\
&&\frac{\Omega^2+mm'g^2}{\omega_{m} \omega_{m'}} \sin(\omega_m t)\sin(\omega_{m'} t),\label{gamma0}\\
&&\gamma_x\equiv \Omega \left[\frac{\sin(\omega_m t)\cos(\omega_{m'} t)}{\omega_{m}}  - \frac{\sin(\omega_{m'} t)\cos(\omega_{m} t)}{\omega_{m'} }\right],\label{gammax}\\
&& \omega_{m}\equiv \sqrt{\Omega^2+m^2g^2}.\label{wm}
\end{eqnarray}
This in particular gives:
\be
\lambda_0=e^{\beta\Omega}.
\ee
The total decoherence factor due to the unobserved environment is then the product of factors (\ref{G_full}) over the traced systems, cf. (\ref{Gamma}), 
\be\label{Gamma_therm}
\Gamma_{mm'}(t)\equiv \prod_{k\in E_{unobs}} \gamma_{mm'}^{(k)}(t).
\ee
We will analyze it below.

We use the same method to calculate the fidelity $F_{mm'}(t) \equiv F(\varrho_m(t),\varrho_{m'}(t))$, with $\varrho_m(t)$ given by (\ref{rhom}, \ref{Um'}) [as before we drop the environment index $k$ for clarity]. Fidelity $F$ can be written as:
\be\label{XX}
F_{mm'}(t)=\frac{1}{Z}\tr\sqrt{X^\dagger X},
\ee
where
\be\label{X}
X\equiv e^{-\beta \Omega S_x} U_{m'}^\dagger U_m e^{-\beta \Omega S_x}. 
\ee
We have to find the eigenvalues of $X^\dagger X$. As before we do it first for $j=1/2$. We find that \cite{NV}:
\begin{eqnarray}
&& \tilde\lambda=\tilde\kappa+\sqrt{\tilde\kappa^2-1},\label{tildelambda}\\
&& \tilde\kappa\equiv \gamma_z^2+\gamma_y^2+\left(\gamma_0^2+\gamma_x^2\right)\cosh(2\beta\Omega)\label{kappa'}\\
&&\gamma_y =-\frac{\Delta m\Omega g}{\omega_{m}\omega_{m'}}\sin(\omega_{m}t)\sin(\omega_{m'}t)\\
&&\gamma_z=g\left[m'\frac{\cos(\omega_{m}t)\sin(\omega_{m'}t)}{\omega_{m'}}-m\frac{\cos(\omega_{m'}t)\sin(\omega_{m}t)}{\omega_{m}}\right]\nonumber\\
&&\\
&& \tilde\lambda_0\equiv e^{2\beta\Omega}=\lambda_0^2,\label{lambda0'}
\end{eqnarray}
and the other eigenvalue is $\tilde\lambda^{-1}$ since $\det(X^\dagger X)=1$. Then:
\be
\tr\sqrt{X^\dagger X}=\sum_{l=-1/2}^{+1/2} \tilde\lambda^l=\frac{\tilde\lambda-\tilde\lambda^{-1}}{\tilde\lambda^{1/2}-\tilde\lambda^{-1/2}}
\ee
For a general $j$, we use the same argument as before when calculating (\ref{G_full}): We extend the above summation  to $l=-j,\dots,j$  and obtain a closed formula for the fidelity:
\be\label{F_full}
F_{mm'}(t) =\frac{\left(\tilde\lambda^{j+1/2}-\tilde\lambda^{-j-1/2}\right)/\left( \tilde\lambda^{1/2}-\tilde\lambda^{-1/2}\right)}{\left(\tilde\lambda_0^{j+1/2}-\tilde\lambda_0^{-j-1/2}\right)/\left( \tilde\lambda_0^{1/2}-\tilde\lambda_0^{-1/2}\right)}.
\ee
Note that now the powers are half of that appearing in (\ref{G_full}), since we were calculating trace of the square root (\ref{XX}).

Looking at the above formulas  (\ref{tildelambda}- \ref{F_full}) it becomes clear why grouping the observed environment into  macrofractions is needed. Due to the finite dimensionality, the single-spin fidelity is periodic in time, so even if it vanishes it does so only momentary. On the other hand, a cumulative
fidelity for a macrofraction according to (\ref{Fmac}) is a product:
\be
F _{mm'}^{mac}(t)=\prod_{k\in mac}F_{mm'}^{(k)}(t) \label{F_therm}.
\ee  
The frequencies $\omega_m^{(k)}$, which govern the time evolution of each factor, can now be randomized \cite{Zurek spins, Mironowicz_PRL, Mironowicz PRA, NV} and the resulting function (\ref{F_therm}) will be quasi periodic in time. If  the macrofraction is big enough, the probability of  $F _{mm'}^{mac}(t)$ returning in the vicinity of its initial value $F _{mm'}^{mac}(0)=1$ will be vanishingly small. Same applies to the decoherence factor  (\ref{Gamma_therm}). One fairly natural way to randomize $\omega_m^{(k)}$  is to assume random coupling constants $g_k$. This corresponds to a situation where we do not fully control the system-environment couplings on the microscopic level.

\section{Thermal environment -- short time behavior} 
Let us analyze short time behavior of (\ref{Gamma_therm}, \ref{F_therm}) to understand better its behavior. We start with the decoherence factor. Expanding $U_{m'}^\dagger U_m$ up to $t^2$ and inserting into (\ref{gmm'}), we obtain for the single-environment the variance formula:
\ben
&&|\gamma_{mm'}(t)| \approx 1-\frac{t^2}{2}\left\langle(H_{m'}-H_m)^2-\langle H_{m'}-H_m\rangle^2\right\rangle\nonumber\\
&&\approx \exp\left[-2t^2\Delta m^2g^2 \left\langle S_z^2-\langle S_z\rangle^2\right\rangle\right],
\een
where we introduced the thermal average $\langle A \rangle \equiv \frac{1}{Z} \tr(A e^{-2\beta \Omega S_x})$ and used (\ref{Hm}).

Calculating the averages, we find that $\langle S_z\rangle=0$ and 
\ben
&&\langle S^2_z\rangle =\frac{1}{2}\left[j(j+1)-\frac{1}{4\Omega^2 Z}\frac{\partial^2Z}{\partial\beta^2}\right]=\\
&&\frac{1}{4}\coth(\beta\Omega)\Big[(2j+1)\coth[(2j+1)\beta\Omega]-\coth(\beta\Omega)\Big].
\een
If the size $\bar N$ of $E_{unobs}$ is large, we can use the law of large numbers (as in the previous studies \cite{Mironowicz_PRL, Mironowicz PRA}) and write the full decoherence factor (\ref{Gamma_therm}) as:
\ben
\left|\Gamma_{mm'}(t)\right|\approx \exp\left[ - 2 \bar N \langle\langle g^2\rangle\rangle \Delta m^2 t^2 \langle S^2_z\rangle\right],
\een
where for simplicity we assumed all $\Omega_k$ being equal and denoted by $\langle\langle \cdot \rangle\rangle$ the average over the distribution of $g_k$. For large-$j$ we may further approximate the above formula by:
\be
\left|\Gamma_{mm'}(t)\right|\approx \exp\left[ - \frac{1}{2} \bar N (2j) \langle\langle g^2\rangle\rangle \Delta m^2 t^2 \coth(\beta\Omega)\right],\label{Gshortfinal}
\ee
from which it is clear that $\bar N$ and $2j$ control the initial decay of coherences in exactly the same way.

Let us now examine the fidleity $F_{mm'}(t)$, expanding it up to $t^2$. We have that (cf. (\ref{XX}, \ref{X})):
\ben
&&F_{mm'}(t)=\tr\sqrt{\sqrt{\varrho_0}U_{m'}^\dagger U_m\varrho_0(U_{m'}^\dagger U_m)^\dagger\sqrt{\varrho_0}},\\
&&U_{m'}^\dagger U_m\varrho_0(U_{m'}^\dagger U_m)^\dagger \approx \varrho_0+it\left[(H_{m'}-H_m),\varrho_0\right].
\een
We then use (\ref{Hm}) and the well know expansion formula for the fidelity  \cite{Braunstein} to obtain:
\ben
&&F_{mm'}(t)\approx\tr\sqrt{\sqrt{\varrho_0}(\varrho_0+\delta \varrho)\sqrt{\varrho_0}} \nonumber\\
&& \approx 1- \frac{1}{4}\sum_{l,l'}\frac{|\delta\varrho_{ll'}|^2}{p_l+p_{l'}}\equiv 1 -\frac{1}{2}g^2\Delta m^2 t^2 \mathcal F 
\een
where $p_k$ are the eigenstates of $\varrho_0$ and $\delta\varrho_{ll'}$ are the matrix elements of $\delta\varrho$ in the eigenbasis of $\varrho_0$, which in our case is that of $S_x$ (cf. (\ref{rho0})). The quantity $\mathcal F$ is the quantum Fisher information (see e.g. \cite{QFI}), which here controls the short-time behavior of the fidelity \cite{gravity}.  It reads:
\be
\mathcal F =2 \sum_{l,l'=-j}^j\frac{(p_l-p_{l'})^2}{p_l+p_{l'}}|\langle l| S_x^2 | l'\rangle|^2,
\ee
where we have rotated the basis to that of $S_z$ so that $S_z \ket l = l \ket l$. Using the fact that $p_l=1/Z e^{-2l\beta\Omega }$, we obtain:
\ben
&&\mathcal F = \nonumber\\
&&2 \sinh^2{\beta\Omega}\left[j(j+1)-\frac{1}{4\Omega^2 Z}\frac{\partial^2Z}{\partial\beta^2}-\frac{\tanh\beta\Omega}{2\Omega}\frac{\partial}{\partial \beta}\log Z\right]\nonumber\\
&&=(2j+1)\tanh(\beta\Omega)\coth[(2j+1)\beta\Omega]-1.
\een
Using the same approximations as in (\ref{Gshortfinal}), we find again that for large-$j$, the macrofraction size $N_{mac}$ and $2j$ control the decay of the fidelity in the same way:
\be
F^{mac}_{mm'}(t)\approx \exp\left[ - \frac{1}{2} N_{mac} (2j) \langle\langle g^2\rangle\rangle \Delta m^2 t^2 \tanh(\beta\Omega)\right]. \label{Fshortfinal}
\ee
The reciprocal behavior w.r.t. the temperature of (\ref{Fshortfinal}) and (\ref{Gshortfinal}) is intuitively clear as hot environments (small $\beta$) decohere more efficiently, but are too noisy to store efficiently information about the central spin. The above results make clear another aspect: The efficiency of information encoding in the environment is exponentially improved with growing $j$.

\begin{figure*}[htbp]
	\begin{center}
		\begin{tabular}{ccc}
			&Sample $g_k$ realization&Average over 100 realizations\\
			\textbf{(a)}&&\\
			&	\includegraphics[width=\columnwidth]{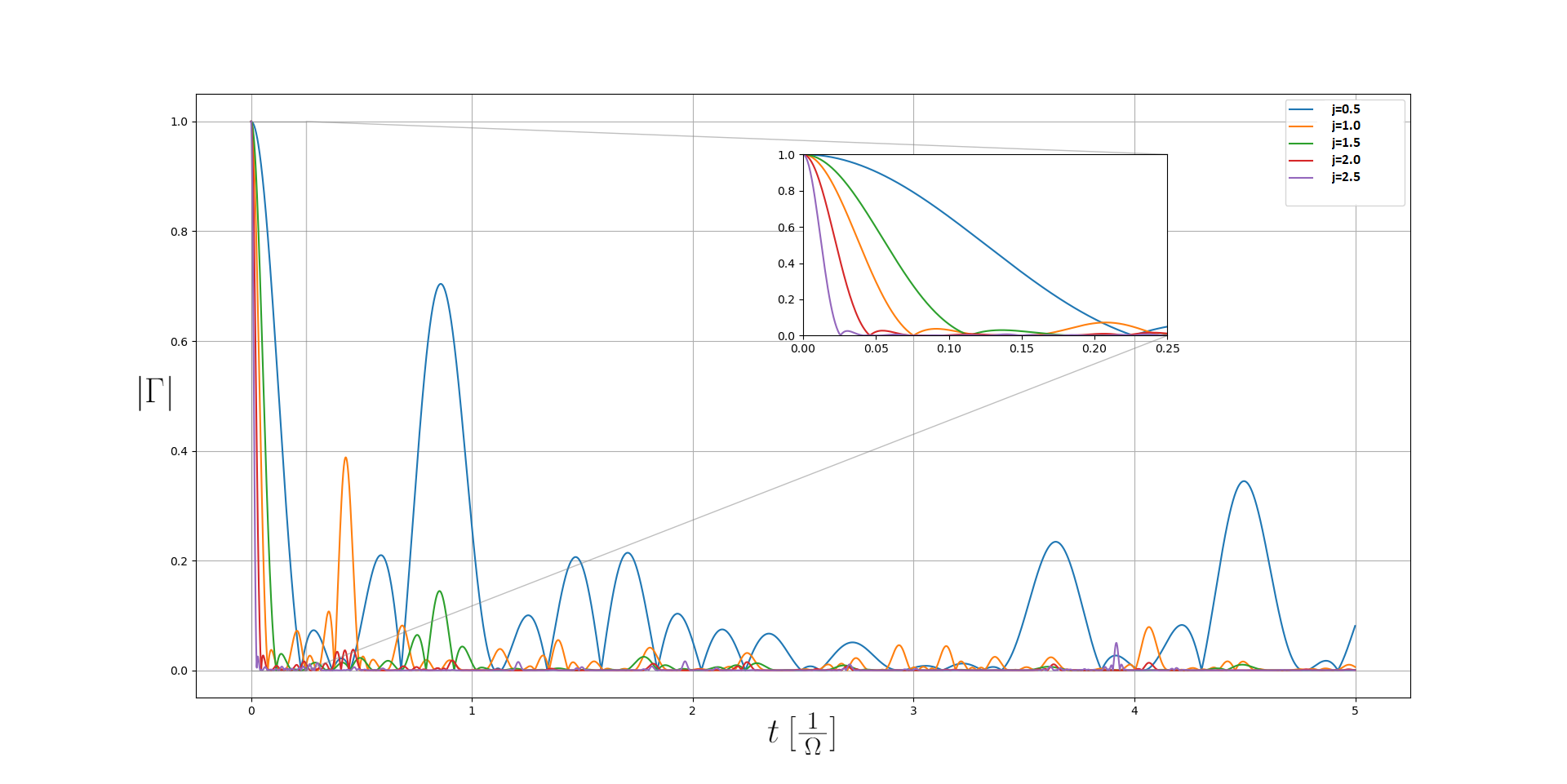}&
			\includegraphics[width=\columnwidth]{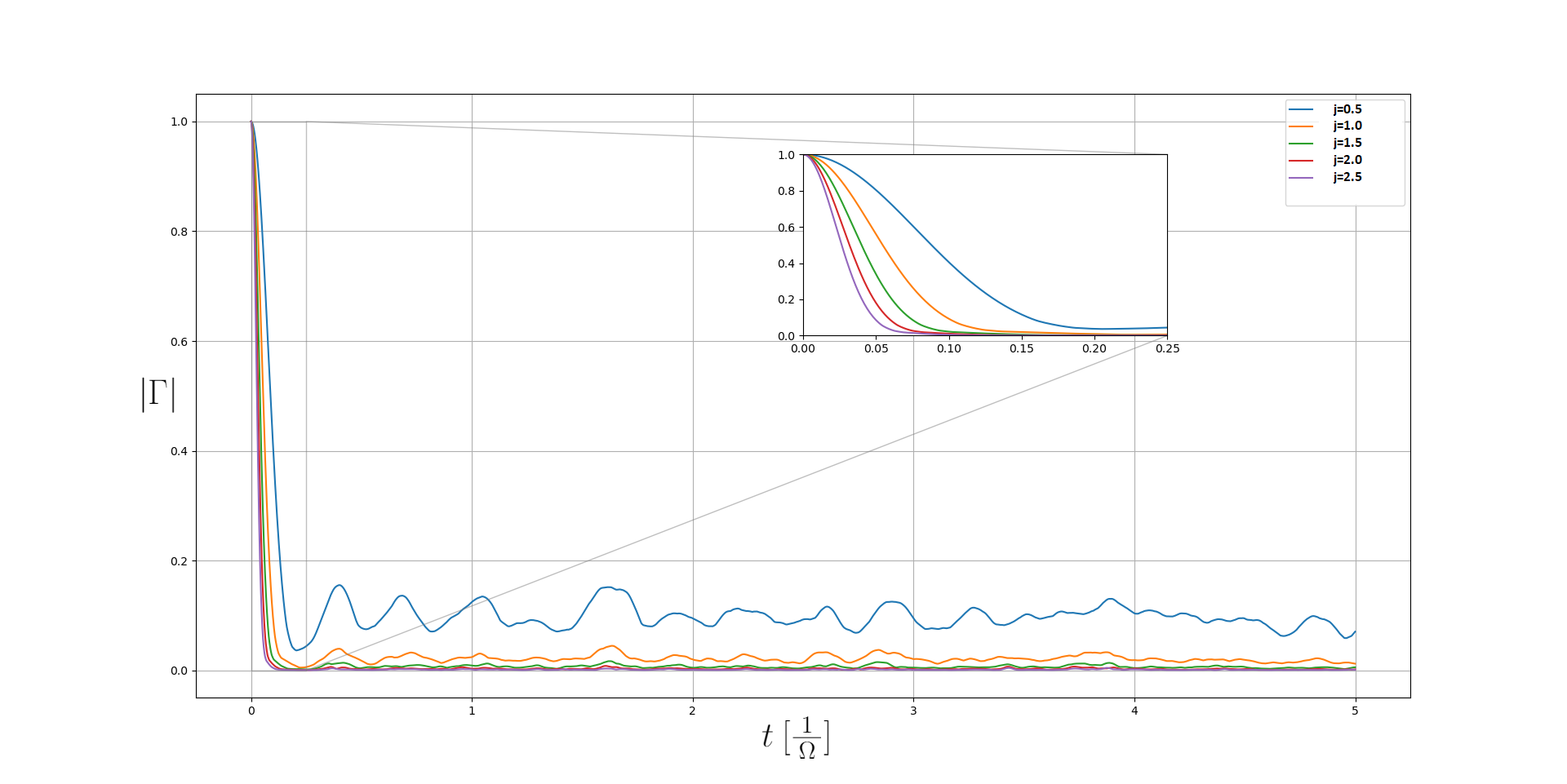}\\
			
			\textbf{(b)}&&\\
			&	\includegraphics[width=\columnwidth]{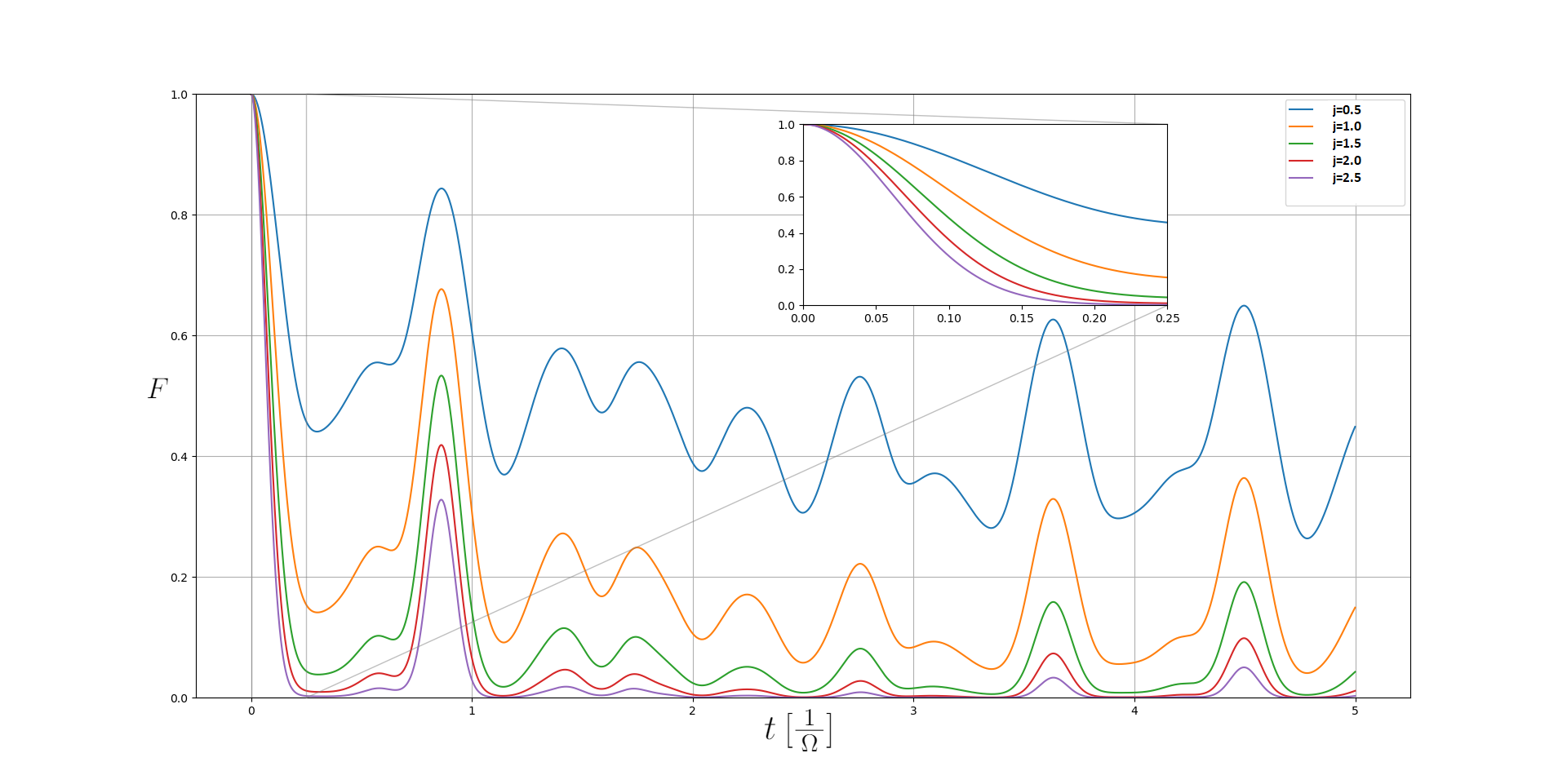}&
			\includegraphics[width=\columnwidth]{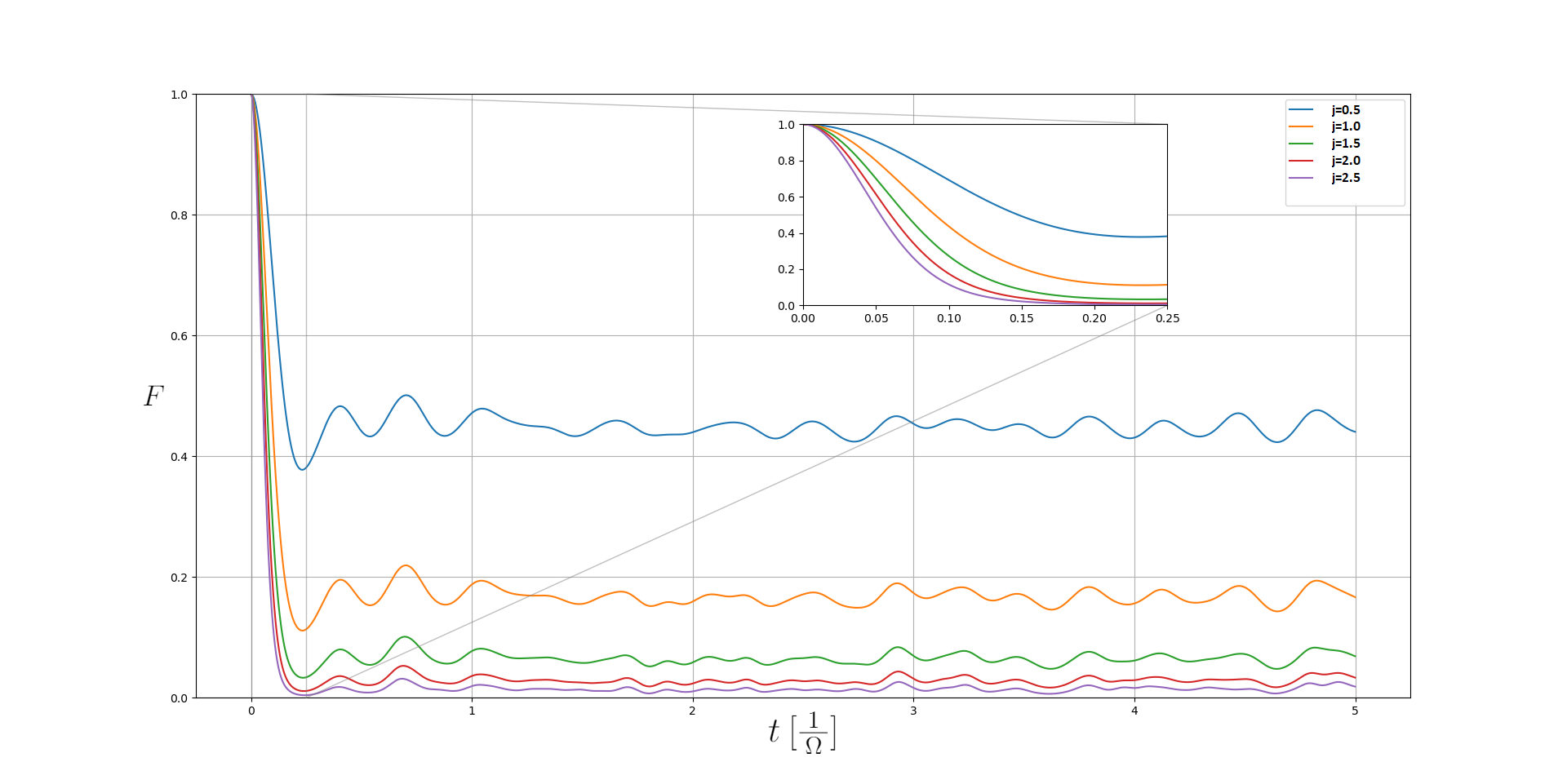}\\
		\end{tabular}
	\end{center}
	\caption{[Color online] Plots of: a) modulus of the decoherence factor (\ref{Gamma_therm}) and b) the fidelity (\ref{F_therm}) as functions of time and the spin $j$. The time units are $\Omega^{-1}$. Left column represents a sample realization of $g_k$, while the right column is the average over $100$ realizations. On all the plots, the blue lines corresponds to $j=1/2$, orange to $j=1$, green to $j=3/2$, red to $j=2$, and violet to $j=5/2$. The rest of the parameters is described in the main text. The inserts show the short time behavior, confirming faster Gaussian decay with the increased spin.}
	\label{fig}
\end{figure*}

\section{Thermal environment -- numerical analysis}
Past short times the analysis of (\ref{Gamma_therm}, \ref{F_therm}), becomes notoriously difficult analytically and we will carry it on numerically. To illustrate the behavior of (\ref{Gamma_therm}, \ref{F_therm}) and SBS formation, we will assume a completely random couplings, drawing $g_k$ from an uniform distribution. Assuming as before the same tunneling frequencies $\Omega_k=\Omega$ for all the environments, it follows from (\ref{kappa}-\ref{wm}) and (\ref{kappa'}-\ref{lambda0'}) that all other parameters scale with $\Omega$. We assume $g_k/\Omega \in [0\dots 10]$ to cover both low and high coupling strength and $\beta\Omega=0.9$, corresponding to the intermediate temperature. We have chosen the latter since: i) for low temperatures, $\beta\Omega \gg 1$,  the initial state is close to a pure one and both decoherence and fidelity factors become identical (cf. (\ref{fid}) which becomes the state overlap for pure states); ii) for high temperatures , $\beta\Omega \approx 0$, the initial state is close to a completely mixed one and although the decoherence is happening, it is not accompanied by distinguishability process -- the states (\ref{rhom}) remain unchanged as the environment is too noisy to store any information about the central spin. The sizes of the unobserved fraction and each of the observed macrofraction are taken to be the same  and $|E_{unobs}|=|mac|=5$. They do not need to be the same, since  these are in general different parts of the environment, but we fixed this sizes for the following reasons: i) in the case of the unobserved part this is a very small fraction and it corresponds to the worst case scenario decoherence-wise; ii) in the case of the observed macrofraction, the chosen small size is close to what is possible to control in experimental systems like NV-centers \cite{ZurekJelezko}.  With the increasing fraction sizes, both decoherence and distinguishability improve due to the larger amount of random phases entering the products (\ref{Gamma_therm}, \ref{F_therm}).  For the central spin we assume two neighboring levels of the lowest value $m=-1/2$, $m'=+1/2$ to introduce only a minimal damping due to the level separation (we recall that the central spin is an independent parameter).  The plots of (\ref{Gamma_therm}, \ref{F_therm}) as a function of time and the environment spin $j$ are presented in Fig. ~\ref{fig}. Left column shows a sample realization, while the right one the average over $100$ realizations of $g_k$.

As we can see, indeed the increase of the spin helps both the decoherence and the distinguishability processes. This is to be expected since spin-$j$ system can be simulated by the symmetric subspace of $2j$ spin-$1/2$ systems, so increasing the spin effectively increases the number of  systems, although without introducing new random phases, which in turn helps the decoherence and the distinguishability processes. This is echoed in the $j$-dependent powers in (\ref{G_full},\ref{F_full}) and in the short-time behavior (\ref{Fshortfinal}, \ref{Gshortfinal}) However, as we can see from Fig. \ref{fig}, the efficiencies of decoherence and distinguishability processes are different. This is especially clear for a sample realization (left column of Fig. \ref{fig}). Although past $j=1/2$ the decoherence is quite efficient, the fidelity function has considerable revivals even for $j=5/2$. This is so because we have deliberately chosen a rather small macrofraction of only $5$ spins, to stay close to what is experimentally realistic, so there is not enough dephasing at all times at the given temperature. However in between the revivals, both the fidelity and decoherence factors are reasonably close to zero already for $j=3/2$, indicating the approach to the SBS state according to (\ref{approach}).  The average, with respect to the coupling constants, behavior of fidelity is more regular and shows a monotonic decay with the growing spin-$j$ (Fig. \ref{fig}, right column).

\section{Conclusions}
We analyzed both decoherence and its more sophisticated version, SBS objectivity, for spin-spin models with arbitrary spins: A central spin $j_S$ interacting via pairwise interactions with a collection of spin-$j$ systems, constituting its environment.  We first derived a completely general, analytic expression (\ref{gamma_general}) for the single-spin decoherence factor in the quantum measurement limit (\ref{H}). It is given in terms of spherical harmonics parameters (\ref{rhoY}) of the initial state of the environment. Next we added Hamiltonian for the environment. Assuming thermal environment, we analyzed objectivization process of central spin state via Spectrum Broadcast Structures (\ref{sbs0}). We derived exact, analytic formulas for both the decoherence factor (\ref{G_full}) and the state fidelity (\ref{F_full}) for arbitrary spin-$j$, generalizing previously obtained results for spin-$1/2$ \cite{Zurek spins, Mironowicz_PRL, Mironowicz PRA, NV}. Further analysis was carried out for short times and then  numerically, confirming that higher spins  make objectivization processes more efficient with all other parameters fixed. This is echoed in the $j$-dependent powers in the exact expressions (\ref{G_full}, \ref{F_full}) and in the exponents of the approximate short-time formulas (\ref{Fshortfinal},\ref{Gshortfinal}). As a result, using higher spins will allow to reduce the portions of the environment that have to be controlled.  We hope this will help in designing and performing future experiments aimed at simulating emergence of objectivity and thus understanding the process better.

\section*{Acknowledgements}
JKK acknowledges the support
by Polish National Science Center (NCN) through the grant
no. 2019/35/B/ST2/01896.


\begin{thebibliography}{99}

\bibitem{decoh}  E. Joos, \emph{et al.}, 
\emph{Decoherence and the Appearancs of a Classical World in Quantum Theory}, 
Springer, Berlin (2003).  M. Schlosshauer, \emph{Decoherence and the Quantum-to-Classical Transition},
Springer, Berlin (2007).

\bibitem{ZurekNature} W. H. Zurek,  Nature Phys. {\bf 5}, 181 (2009). 

\bibitem{PRA} R. Horodecki, J. K. Korbicz, and P. Horodecki, Phys. Rev. A {\bf 91}, 032122 (2015). 

\bibitem{Le} T. P. Le and A. Olaya-Castro
Phys. Rev. Lett. {\bf 122}, 010403 (2019).

\bibitem{PRL}   J. K. Korbicz, P. Horodecki, and R. Horodecki, Phys. Rev. Lett. {\bf 112}, 120402 (2014). 

\bibitem{solo} J. K. Korbicz,  arXiv:2007.04276 (2020). 

\bibitem{Mironowicz_PRL}  P. Mironowicz, J. K. Korbicz, and P. Horodecki, Phys. Rev. Lett. {\bf 118}, 150501 (2017). 

\bibitem{Tuziemski qbm}  J. Tuziemski and J. K. Korbicz, EPL {\bf 112}, 40008 (2015).

\bibitem{Lewenstein} A. Lampo, J. Tuziemski, M. Lewenstein, and J. K. Korbicz,  Phys. Rev. A 96, 012120 (2017).

\bibitem{QED}  J. Tuziemski, P. Witas, J. K. Korbicz, Phys. Rev. A 97, 012110 (2018).  

\bibitem{gravity}  J. K. Korbicz and J. Tuziemski,  Gen. Relativ. Gravit. 49:152 (2017). 

\bibitem{measurements}  J. K. Korbicz, E. A. Aguilar, P. \'Cwikli\'nski, and P. Horodecki, Phys. Rev. A {\bf 96}, 032124 (2017). 

\bibitem{GPT} C. M. Scandolo, R. Salazar, J. K. Korbicz, and P. Horodecki, arXiv:1811.01808 (2018).

\bibitem{Kasia}  K. Roszak and J. K. Korbicz, Phys. Rev. A {\bf 100}, 062127 (2019). 

\bibitem{Mironowicz PRA} P. Mironowicz, P. Nale\.zyty, P. Horodecki, and J. K. Korbicz, Phys. Rev. A {\bf 98}, 022124 (2018).

\bibitem{NV}  D. Kwiatkowski, \L. Cywi\'nski, J. K. Korbicz, New J. Phys. {\bf 23}, 043036 (2021). 

\bibitem{Zurek spins} F. M. Cucchietti, J. P. Paz, and W. H. Zurek, \emph{Decoherence from spin environments}, Phys. Rev. A {\bf 72}, 052113 (2005).

\bibitem{Mauro} E. Ryan, M. Paternostro, and S. Campbell,  	arXiv:2011.13385 (2020). 

\bibitem{magnets} M. N. Leuenberger and D. Loss, Nature {\bf 410}, 789 (2001).

\bibitem{Hamdouni} Y. Hamdouni, Phys. Rev. A {\bf 94}, 022120 (2016). 

\bibitem{Ciampini} M. A. Ciampini, G. Pinna, P. Mataloni, and M. Paternostro, Phys. Rev. A {\bf 98}, 020101 (2018).

\bibitem{Chen} M.-C. Chen, H.-S. Zhong, Y. Li, D. Wu, X.-L. Wang, L. Li, N.-L. Liu, C.-Y. Lu, and J.-W. Pan, Sci. Bull. {\bf 64}, 580 (2019).

\bibitem{ZurekJelezko} T. Unden, D. Louzon, M. Zwolak, W. H. Zurek, and F. Jelezko, Phys. Rev. Lett. {\bf 123}, 140402 (2019).

\bibitem{Fuchs} C. A. Fuchs, J. van de Graaf, IEEE Trans. on Inf. Theor. {\bf 45}, 1216 (1999). 

\bibitem{Perelomov} A. Perelomov, \emph{Generalized Coherent States and Their Applications}, Springer, Berlin (1986).

\bibitem{Arecchi} F. T. Arecchi, E. Courtens, R. Gilmore, and H. Thomas, Phys. Rev. A {\bf 6}, 2211, (1972).

\bibitem{Braunstein} S. L. Braunstein and C. M. Caves, Phys. Rev. Lett. {\bf 72}, 3439 (1994).

\bibitem{QFI}  G. T\'oth and I. Apellaniz, J. Phys. A: Math. Theor. {\bf 47}, 424006 (2014). 


\end{thebibliography}
\end{document}